\documentclass[fleqn,twoside]{article}
\usepackage{espcrc2}

\usepackage{graphicx}

\newcommand{\AmS}{{\protect\the\textfont2
  A\kern-.1667em\lower.5ex\hbox{M}\kern-.125emS}}

\newcommand{\Bsl}{{B\hspace{-5pt}{/}}}
\newcommand{\ndop}{{\mathcal O}_{R/L;L}^{\Bsl}}
\newcommand{\msbar}{\overline{\mbox{MS}}}
\newcommand{\Prt}[1]{{\mathcal P}^{#1}}
\newcommand{\Swt}[1]{{\mathcal S}^{#1}}

\title{Nucleon Decay with Domain-Wall Fermions}

\author{Yasumichi Aoki\address{RIKEN BNL Research Center, Brookhaven
National Laboratory, Upton, NY 11973, USA}\address{Physics Department,
Columbia University, New York, NY 10027, USA} \thanks{Present address:
Physics Department, University of Wuppertal, 
42097 Wuppertal, Germany. }
[RBC collaboration]
\thanks{We thank RIKEN, Brookhaven National Laboratory
and the U.S.\ Department of Energy for providing the facilities
essential for the completion of this work.}}

\begin{document}

\begin{abstract}
 We report on our on-going project to calculate the nucleon decay matrix
 elements with domain-wall fermions.
 Operator mixing is discussed employing a non-perturbative renormalization.
 Bare matrix elements of all the possible decay modes induced by the
 dimension-six operators are calculated with the direct method, which
 are compared with the indirect calculation using chiral perturbation
 theory. 
\end{abstract}

\maketitle

\section{INTRODUCTION}

There are two methods to calculate nucleon decay matrix elements using
lattice QCD. One is the direct calculation of the 
matrix elements dealing with three- and various two-point functions
\cite{Gavela:1989cp,Aoki:1999tw,Aoki:2002ji}. 
The other is the indirect method where one relies on chiral perturbation 
theory whose low energy constants are calculated on the lattice 
\cite{Hara:1986hk,Bowler:1988us,Gavela:1989cp,Aoki:1999tw,Aoki:2002ji}.
The former 
method is preferable but requires huge computation, typically 10
times larger than the latter to achieve similar statistical
accuracy. 

In a previous work \cite{Aoki:2002ji}, we made use of good chiral
symmetry of the domain-wall fermion with the DBW2 gauge action
\cite{Aoki:2002vt}.  
The proton decay matrix element of the pion final state was calculated 
with both methods. The results are consistent with each other within 
the relatively large error. There we assumed the mixing of operators 
arising from the explicit chiral symmetry breaking is negligible. 
Here we reconsider this issue by calculating the mixing with a 
non-perturbative renormalization (NPR).
Also the matrix elements of the proton decay in case
of the non-degenerate quark mass in the pseudoscalar state are 
calculated allowing further discussion on the effectiveness of the
chiral perturbation theory. 

All of our calculations are done on the quenched DBW2 gauge configurations
with size $N_\sigma^3\times N_\tau=16^3\times 32$. The domain-wall
height is $M=1.8$, the fifth dimension size is $L_s=16$ for the NPR and
$12$ for the matrix element.

\section{OPERATOR MIXING}

The nucleon decay is induced by the dimension-six operator which
consists of one lepton and three quarks. The three quark operator,
with which we calculate matrix element with initial nucleon and final
pseudoscalar state, in general takes a form as
\begin{equation}
\mathcal{O}^{a}_{uds}= \epsilon^{ijk}(u^{iT}
C \Gamma d^j) \Gamma' s^k.
\end{equation}
The superscript $a$ specifies a combination of the gamma matrix 
$(\Gamma \Gamma')$ listed in Table \ref{tab:mix}.
Operators are classified by the ordinary parity ($\Prt{}$)
and by the parity of switching ($\Swt{}$) $u$ and $d$ quarks.
The $\Swt{}$ is a symmetry of the (lattice) QCD Lagrangian when quark
masses are degenerate. 
The $u$, $d$, $s$ do not necessarily mean the real flavor, but simply
stand for the different flavors. The other arrangements of flavors,
such like ${\mathcal O}_{usd}^a$ and ${\mathcal O}_{dus}^a$ are always
expressed in terms of ${\mathcal O}_{uds}^a$ by Fierz rearrangement and/or
by $\Swt{}$. Operators with two flavors like  ${\mathcal O}_{udu}$ 
follows the same argument.
\begin{table}[t]
 \caption{Operator classification by parity and switching symmetry.}
 \label{tab:mix}
 \begin{center}
  \begin{tabular}{|c||ccc|cc|}
   \hline
   & & $\Swt{-}$ & & \multicolumn{2}{c|}{$\Swt{+}$} \\
   \hline
   \hline
   $\Prt{+}$ & $SS$ & $PP$ & $AA$ & $VV$ & $TT$\\
   \hline
   $\Prt{-}$ & $SP$ & $PS$ & $AV$ & $VA$ & $T\tilde{T}$\\
   \hline
  \end{tabular}
 \end{center}
 \vspace{-12pt}
\end{table}
Lattice operators are renormalized as 
\begin{equation}
{\mathcal O}_{ren}^a = Z^{ab} {\mathcal O}_{latt}^b.
\label{eq:mix}
\end{equation}
For example, for operators in $\Prt{+}$, $\Swt{-}$
sector, which will not mix the operators in the other sectors,
$a$ and $b$ in eq.~\ref{eq:mix} may be $SS$, $PP$, $AA$.
The same $3\times 3$ matrix $Z$ applies to the $\Prt{-}$, $\Swt{-}$
sector: $SP$, $PS$, $-AV$, too.

For the NPR with RI-MOM scheme \cite{Martinelli:1995ty,Blum:2001sr}
we calculate the Green's function of the operator with the three quark
external states in the Landau gauge,
\begin{equation}
 G^{a}(x_0,x_1,x_2,x_3)=\langle {\mathcal O}_{uds}^{a}(x_0)
 \bar{u}(x_1) \bar{d}(x_2) \bar{s}(x_3) \rangle.
\end{equation}
We set $x_0=0$. Fourier transformation is performed on the other
legs with the same momentum $p$, which are then amputated 
to obtain the vertex function,
\begin{equation}
 \Lambda^{a}(p^2) = \mbox{F.T.}\ G^{a}(0,x_1,x_2,x_3) |_{Amp}.
\end{equation}
Writing the tensor indices explicitly, the renormalization condition of
the RI-MOM scheme leads, 
\begin{equation}
 P^a_{ijk\;\beta\alpha\;\delta\gamma} \cdot
  Z_q^{-3/2} Z^{bc} \Lambda^c_{ijk\;\alpha\beta\;\gamma\delta}
  = \delta^{ab},
\end{equation}
where $Z_q$ is the quark wave function renormalization $i, j, k$ are 
color indices, $\alpha$, $\beta$ and $\gamma$, $\delta$ 
are Dirac indices associated with $\Gamma$, $\Gamma'$ respectively. 
The renormalization condition should be applied where the 
pion pole effect \cite{Blum:2001sr} is absent and the lattice  
artifact is small : $\Lambda_{QCD}\ll |p| \ll 1/a$. 
The projection matrix $P^a$, for example onto $SS$, is
$P^{SS}=\epsilon^{ijk}(C^{-1})^{\beta\alpha}
\delta^{\delta\gamma}/(6\cdot 4\cdot 4)$.
We define the matrix $M$ as,
\begin{equation}
 M^{ab} = P^a_{ijk\;\beta\alpha\;\delta\gamma} \cdot
 \Lambda^b_{ijk\;\alpha\beta\;\gamma\delta}, 
\end{equation}
which will give $Z_q^{3/2}(Z^{-1})^{ab}$.

\begin{figure}[t]
 \begin{center}
  \includegraphics[width=6.8cm]{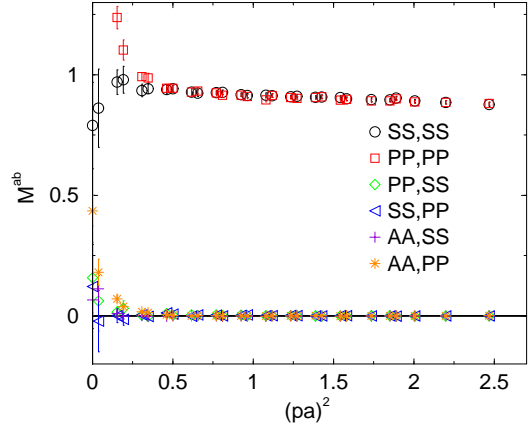}
 \end{center}
 \vspace{-32pt}
 \caption{Elements of $M$ for $m_f=0.025$ as a function of momentum squared.}
 \label{fig:mix1}
 \vspace{-12pt}
\end{figure}

Fig.~\ref{fig:mix1} shows the resulting $M^{ab}$
at a quark mass $m_f=0.025$ for the domain-wall fermion.
51 configurations have been analyzed \cite{Dawson:2002nr}.
The renormalization factor of $SS,SS$ and $PP,PP$ 
are identical within the error except for the vicinity of 
zero momentum, the others are zero within
the two standard deviation level. Absence of the off-diagonal element and
coincidence of the $SS,SS$ and $PP,PP$ element are expected in the case of
exact chiral symmetry. 
We have examined four different masses $m_f=0.025, 0.04, 0.055, 0.07$,
where mass dependences are hardly seen in all range of momentum except
near zero.

As a result, the nucleon decay operator 
$\ndop=\epsilon^{ijk}(u^{iT} CP_{R/L}d^j)P_{L}s^k$ [$S\pm P\ S-P$] is
renormalized  
multiplicatively without any mixing at practical level for this 
parameter of the domain-wall fermion. 
In the next section we show the results of the bare matrix elements 
with $L_s=12$. The residual mass $m_{res} = 1.6$ MeV at $L_s=12$ is
still comparable to $0.7$ MeV at $L_s=16$ \cite{Aoki:2002vt}.
Hence the effect of $L_s$ on the renormalization factor
is expected to be absent in this case, too, which needs to be checked 
by further performing simulation with $L_s=12$. 

\section{MATRIX ELEMENTS}

\begin{figure}[t]
   \begin{center}
    \includegraphics[width=7.6cm]{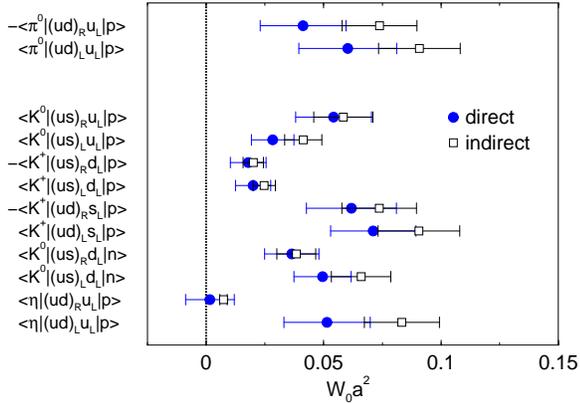}
   \end{center}
 \vspace{-32pt}
 \caption{Summary $W_0$ for both direct and indirect calculations with
 unrenormalized operators.}
\label{fig:W0}
 \vspace{-16pt}
\end{figure}
We follow the ref.~\cite{Aoki:1999tw} to calculate the nucleon
decay matrix elements by the direct method. Our results for the
decay to a meson with degenerate quark mass are already shown in
\cite{Aoki:2002ji}, where we have used 100 gauge configurations.
We extend our calculation to the non-degenerate quark mass 
case to treat the decay to kaon.
Our masses are $m_{f1}, m_{f2} = 0.02, 0,04, 0.06, 0.08$ with 
$m_{f1}\le m_{f2}$, where $m_{f1}$ is for the $u$, $d$ quarks, $m_{f2}$
for  the $u$, $d$ or $s$ quark.
The other parameters are the same as in \cite{Aoki:2002ji}. 
Result of the relevant form factor $W_0$ of the matrix element for various
decay modes are shown in Fig.~\ref{fig:W0}.
We are assuming the SU$(2)$ symmetry for the 
$u$ and $d$ quarks. There are other possible matrix elements, but
they can be calculated with the matrix elements in the figure
when the SU$(2)$ symmetry is intact.

The indirect method uses the chiral perturbation theory, which relates 
the low energy parameters in the chiral Lagrangian to the matrix
elements. JLQCD extended the calculation by Claudson et.~al
\cite{Claudson:1982gh} to all possible decay modes \cite{Aoki:1999tw}. 
The low energy parameters are mass ratio of nucleon and baryon with
strangeness $S=-1$, $m_N/m_B$, pion decay constant $f$, $D$ and $F$ 
parameterizing the semileptonic decay of baryons, nucleon decay parameter
$\alpha$ and $\beta$. These parameters are in principle calculable on
the lattice. As we are using the chiral fermion, observables calculated
on the lattice should obey the chiral perturbation whose parameters
are calculated on the lattice. However we use the experimental values
except for $\alpha$ and $\beta$. The reasons are that we know the
results of $f_\pi$ \cite{Aoki:2002vt} and $g_A$ \cite{Sasaki:2003jh}
$(=D+F)$ are anyway consistent with the experiment, and that
we have no estimate of $m_B$ and individual $D$ and $F$.
Using the $\alpha$ and $\beta$ in our previous study \cite{Aoki:2002ji} and
$D=0.8, F=0.47$ \cite{Hsueh:1988ar}, $f=0.131$ GeV, $m_N/m_B=0.817$,
the matrix elements are calculated and shown in Fig.~\ref{fig:W0}.
The lattice cutoff calculated by $\rho$ mass input, $a^{-1}=1.31$ GeV is
used to make dimensionless value. 

The indirect method gives quite good estimate of the matrix element.
Consistency of the two method within the error is in contrast to the 
result of JLQCD \cite{Aoki:1999tw} with Wilson fermion. However, this
could be caused by larger statistical 
error in our calculation. We need more statistics to be able to make
more reliable statements.

\section{SUMMARY}
We have investigated the mixing of three-quark nucleon decay operators 
with domain-wall fermions by NPR. The mixing is practically negligible
at the parameter set similar to that where the matrix element are
calculated. We have extended our nucleon decay matrix element to those
for the final kaon state matrix elements. The direct and indirect
methods give consistent results within our statistical error. 
Performing NPR at the parameter of matrix element calculation 
together with the matching calculation will give us the matrix element
in $\msbar$.

\bibliography{proc}
\bibliographystyle{h-elsevier}

\end{document}